\PassOptionsToPackage{unicode}{hyperref}
\PassOptionsToPackage{hyphens}{url}
\documentclass[11pt,a4paper]{article}
\usepackage{amsmath,amssymb}
\usepackage[T1]{fontenc}
\usepackage[utf8]{inputenc}
\usepackage{lmodern}

\usepackage{fancyvrb}

\DefineVerbatimEnvironment{Highlighting}{Verbatim}{commandchars=\\\{\}}
\newenvironment{Shaded}{}{}

\newcommand{\AttributeTok}[1]{\textcolor[rgb]{0.49,0.56,0.16}{#1}}

\newcommand{\BuiltInTok}[1]{\textcolor[rgb]{0.00,0.50,0.00}{#1}}

\newcommand{\CommentTok}[1]{\textcolor[rgb]{0.38,0.63,0.69}{\textit{#1}}}

\newcommand{\ControlFlowTok}[1]{\textcolor[rgb]{0.00,0.44,0.13}{\textbf{#1}}}

\newcommand{\DecValTok}[1]{\textcolor[rgb]{0.25,0.63,0.44}{#1}}

\newcommand{\FloatTok}[1]{\textcolor[rgb]{0.25,0.63,0.44}{#1}}

\newcommand{\ImportTok}[1]{\textcolor[rgb]{0.00,0.50,0.00}{\textbf{#1}}}

\newcommand{\KeywordTok}[1]{\textcolor[rgb]{0.00,0.44,0.13}{\textbf{#1}}}
\newcommand{\NormalTok}[1]{#1}
\newcommand{\OperatorTok}[1]{\textcolor[rgb]{0.40,0.40,0.40}{#1}}

\newcommand{\SpecialCharTok}[1]{\textcolor[rgb]{0.25,0.44,0.63}{#1}}
\newcommand{\SpecialStringTok}[1]{\textcolor[rgb]{0.73,0.40,0.53}{#1}}
\newcommand{\StringTok}[1]{\textcolor[rgb]{0.25,0.44,0.63}{#1}}
\newcommand{\VariableTok}[1]{\textcolor[rgb]{0.10,0.09,0.49}{#1}}

\usepackage{longtable,booktabs,array}
\usepackage{calc} 
\usepackage{etoolbox}
\usepackage{jcappub}

\DeclareUnicodeCharacter{21A9}{\ensuremath{\hookleftarrow}}

\usepackage[]{natbib}
\begin{document}
\title{A Simulation Framework for the \textit{LiteBIRD} Instruments}

\makeatletter
\def\@xfootnote[#1]{%
  \protected@xdef\@thefnmark{#1}%
  \@footnotemark\@footnotetext}
\makeatother
\author[1,2]{M.\,Tomasi,\footnote[*]{Corresponding author.}}
\author[3,4,5]{L.\,Pagano,}
\author[6]{A.\,Anand,}
\author[7,8,9]{C.\,Baccigalupi,}
\author[10]{A.\,J.\,Banday,}
\author[3,4]{M.\,Bortolami,}
\author[3,6]{G.\,Galloni,}
\author[11]{M.\,Galloway,}
\author[12,13]{T.\,Ghigna,}
\author[14]{S.\,Giardiello,}
% ...
\affiliation[1]{Dipartimento di Fisica, Università degli Studi di Milano, Via Celoria 16 - 20133, Milano, Italy}
\affiliation[2]{INFN Sezione di Milano, Via Celoria 16 - 20133, Milano, Italy}
\affiliation[3]{Dipartimento di Fisica e Scienze della Terra, Università di Ferrara, Via Saragat 1, 44122 Ferrara, Italy}
\affiliation[4]{INFN Sezione di Ferrara, Via Saragat 1, 44122 Ferrara, Italy}
\affiliation[5]{Université Paris-Saclay, CNRS, Institut d’Astrophysique Spatiale, 91405, Orsay, France}
\affiliation[6]{Dipartimento di Fisica, Università di Roma Tor Vergata, Via della Ricerca Scientifica, 1, 00133, Roma, Italy}
\affiliation[7]{International School for Advanced Studies (SISSA), Via Bonomea 265, 34136, Trieste, Italy}
\affiliation[8]{INFN Sezione di Trieste, via Valerio 2, 34127 Trieste, Italy}
\affiliation[9]{IFPU, Via Beirut, 2, 34151 Grignano, Trieste, Italy}
\affiliation[10]{IRAP, Université de Toulouse, CNRS, CNES, UPS, Toulouse, France}
\affiliation[11]{Institute of Theoretical Astrophysics, University of Oslo, Blindern, Oslo, Norway}
\affiliation[12]{International Center for Quantum-field Measurement Systems for Studies of the Universe and Particles (QUP), High Energy Accelerator Research Organization (KEK), Tsukuba, Ibaraki 305-0801, Japan}
\affiliation[13]{Kavli Institute for the Physics and Mathematics of the Universe (Kavli IPMU, WPI), UTIAS, The University of Tokyo, Kashiwa, Chiba 277-8583, Japan}
\affiliation[14]{School of Physics and Astronomy, Cardiff University, Cardiff CF24 3AA, UK}
% ...

\emailAdd{maurizio.tomasi@unimi.it}

\abstract{\textit{LiteBIRD}, the Lite (Light) satellite for the study of $B$-mode polarization and Inflation from cosmic background Radiation Detection, is a space mission focused on primordial cosmology and fundamental physics.

In this paper, we present the \textit{LiteBIRD} Simulation Framework (LBS), a Python package designed for the implementation of pipelines that model the outputs of the data acquisition process from the three instruments on the \textit{LiteBIRD} spacecraft: LFT (Low-Frequency Telescope), MFT (Mid-Frequency Telescope), and HFT (High-Frequency Telescope). LBS provides several modules to simulate the scanning strategy of the telescopes, the measurement of realistic polarized radiation coming from the sky (including the Cosmic Microwave Background itself, the Solar and Kinematic dipole, and the diffuse foregrounds emitted by the Galaxy), the generation of instrumental noise and the effect of systematic errors, like pointing wobbling, non-idealities in the Half-Wave Plate, \emph{et cetera}.

Additionally, we present the implementation of a simple but complete pipeline that showcases the main features of LBS. We also discuss how we ensured that LBS lets people develop pipelines whose results are accurate and reproducible.

A full end-to-end pipeline has been developed using LBS to characterize the scientific performance of the \textit{LiteBIRD} experiment. This pipeline and the results of the first simulation run are presented in Puglisi et al.~(2025).}
\maketitle

\section{Introduction}\label{introduction}

\textit{LiteBIRD}---the Lite (Light) satellite fo the study of $B$-mode polarization and Inflation from cosmic background Radiation Detection---is a large-class satellite mission proposed by the Institute of Space and Astronautical Science (ISAS), Japan Aerospace Exploration Agency (JAXA), whose launch is planned for the 2030s. It will observe the full sky in 15 frequency bands from 34 to 448 GHz for 3 years, with effective polarization sensitivity of \(2.2\ \mu \mathrm{K\cdot arcmin}\) and angular resolution of 31 arcmin (at 140\,GHz), employing 4508 detectors sampling at 19.1\,Hz \citep{Hazumi2021PTEP}. Its main goal is to observe the polarization of the Cosmic Microwave Background (CMB) radiation, with the aim of testing the validity of the inflationary paradigm. The target sensitivity of \textit{LiteBIRD} is \(\delta r \leq 10^{-3}\), where \(\delta r\) is the uncertainty of the tensor-to-scalar ratio, \(r\). This sensitivity will let us test major single-field inflation models.

\textit{LiteBIRD} will host three instruments onboard the spacecraft: the Low-Frequency Telescope (LFT), the Mid-Frequency Telescope (MFT), and the High-Frequency Telescope (HFT). All the instruments employ Transition-Edge Sensor (TES) bolometers to measure the intensity and the polarization of the radiation from the sky, which will be focused on the focal planes of each instrument employing refractive and reflective telescopes.

To validate the design of the instruments onboard the spacecraft, the \textit{LiteBIRD} collaboration has identified the need for a framework, called LBS, that provides a set of modules to model several aspects of the data acquisition process of the instruments, including the most relevant systematic effects. This framework is a Python library the collaboration has used to develop an end-to-end (E2E) \emph{simulation pipeline} that is described in \cite{puglisi2024}; the purpose of the pipeline is to simulate the production of the time-ordered data that will be acquired by the actual instrument once deployed in space, to determine whether the design of the experiment can achieve its scientific objectives. Another important purpose of this pipeline is to produce output data that can be used as input by data-reduction pipelines; this is the case of the work described in \cite{aurvik2024}, where the authors process the output of the E2E simulations using \texttt{Commander3} to estimate the amount of resources needed to perform a Bayesian end-to-end analysis of the \textit{LiteBIRD} data. In this work, we describe the framework itself.

We decided to base the simulation pipeline on a framework instead of coding the whole E2E pipeline directly because this ensures a few advantages:

\begin{itemize}
\item The design phase of \textit{LiteBIRD} needs several pipelines: apart from the E2E pipeline, the team requires specific pipelines to simulate targeted effects like Half-Wave Plate (HWP) systematics or the observation of transient sources, as well as simpler pipelines that produce approximated results in a fraction of the time needed by full simulations. Implementing a common framework speeds up the development, because each pipeline can be built by joining several ready-made building blocks.
\item Reusing the same framework for many pipelines ensures consistency between them, particularly concerning the mathematical models used to describe the hardware and the file formats used to load input data and to save the results of the simulations.
\end{itemize}

The structure of this work is the following. In Section~\ref{sec-requirements} we describe the requirements that have driven the implementation of LBS, including the memory layout and the need to properly track the provenance of the inputs to ensure that the results produced using LBS are reproducible. In Section~\ref{sec-design} we show how LBS was implemented and what are its main characteristics. Section~\ref{sec-modules} provides a description of the most important modules that are available in LBS 0.11.0, the version used to run the E2E tests described in \cite{puglisi2024}. We explain how we validated the implementation of LBS in Section~\ref{sec-validation}. The last part of the paper, Section~\ref{sec-full-example}, provides the full implementation of a simple pipeline and showcases the main features of LBS that have been presented in the previous sections.

\section{Requirements}\label{sec-requirements}

The \textit{LiteBIRD} Simulation Team was established in 2019 to develop a simulation framework for the collaboration. The following core requirements were established since the beginning:

\begin{itemize}
\item \textbf{Development of pipelines}. The framework must facilitate the development of simulation pipelines that can generate realistic timelines and maps. These outputs are crucial for validating the design of the scientific instruments and the overall mission architecture.
\item \textbf{Computational efficiency}. Due to the limited computational resources available on high-performance computing (HPC) clusters, the framework must exhibit high performance in terms of memory utilization and execution time.
\item \textbf{Availability of fundamental simulation components}. The framework must provide a comprehensive set of building blocks that can be assembled to build the simulation pipelines needed by the \textit{LiteBIRD} collaboration.
\item \textbf{Language Familiarity}. The framework must be implemented in a programming language that is widely adopted within the collaboration, as this minimizes the barrier to entry for new developers.
\item \textbf{Ease of contribution}. The architecture of the framework and the development practices must be structured to enable reasonably straightforward contributions from collaboration members.
\item \textbf{Comprehensive documentation}. This ensures maintainability and enables users to utilize the framework's capabilities fully.
\item \textbf{Reproducibility of results}. The framework must incorporate tools and mechanisms that enhance the ability to reproduce simulation results.
\item \textbf{Integration with an Instrument Model database (IMo)}. The framework must use a well-established way to retrieve its inputs from a common database containing a description of the instrument, which is handled by the \textit{LiteBIRD} Instrument MOdel team (IMo team).
\end{itemize}

We now define the meaning of an \emph{end-to-end (E2E) pipeline}, because it helps to understand the design of our framework better and puts this work in the context of what is described in \cite{puglisi2024}.

The main scientific goal of \textit{LiteBIRD} is to set an upper limit to the value \(\delta r\), the uncertainty on the scalar-to-ratio parameter \(r\), under the assumption of a fiducial model with $r = 0$ \cite{Hazumi2021PTEP}. Thus, E2E simulations should go through the following steps:

\begin{enumerate}
\item Start from a reasonable estimate of the sky signal;
\item Model the data acquisition process of the detectors, considering the way the instruments are built and mounted within the optical system, the movements of the spacecraft, and other details of the mission;
\item The result of the simulation of data acquisition is a set of timelines, which are used to produce sky maps at different frequencies in the range 34--448\,GHz;
\item Combine and process the sky maps using component-separation codes to produce estimates of the CMB as well as other signals (dust, synchrotron emission, etc.);
\item Compute power spectra from the CMB map;
\item Estimate \(r\) from the timelines, maps, and power spectra.
\end{enumerate}

This list prompts us to make two crucial remarks. First, the data acquisition process is modeled only in the first \emph{two steps} of the procedure: technically speaking, steps 3--6 are \emph{data reduction tasks}, which are the same for simulated and real data. However, to ensure that the results of E2E simulations can be accurately interpreted and fed into the next stages of the analysis, we had to include a few data-reduction modules (map-makers) in our simulation framework, as explained in Section~\ref{sec-map-making}. The second remark is that a certain number \(N\) of Monte Carlo realizations is required to estimate error bars for the scientific parameters produced by E2E simulations. This increases the processing power required to run the simulations, necessitating the framework to be optimized for both speed and memory.

E2E simulations are not the only kind of simulations needed for an experiment of the scale of \textit{LiteBIRD}. For instance, to characterize the ability to perform in-flight calibrations, the \textit{LiteBIRD} collaboration has developed dedicated pipelines to simulate the observation of bright objects (planets). Thus, the list of modules to be included in the framework is not exhausted once all the modules needed in a E2E pipeline simulating nominal operations are implemented.

\section{Overall design}\label{sec-design}

Now that we have listed the requirements, in this section we present the architectural design of LBS and describe the elements that enable the simulation of the three instruments onboard the \textit{LiteBIRD} spacecraft. The features listed in this section are used by all the simulation modules provided by LBS (see Section~\ref{sec-modules}).

\subsection{Supported platforms}\label{sec-supported-platforms}

LBS is implemented in Python and can be used on 64-bit Unix machines\footnote{We are not able to support Windows systems natively because our code relies on Healpy, which is currently not available under this operating system (see \url{https://github.com/healpy/healpy/issues/25}). One can however install LBS within the Windows Subsystem for Linux (WSL).} like Linux and Mac OS X. In each release, we support Python versions whose End of Life (EoL) is more than one year in the future and that are supported by NumPy \citep{harris2020array} and Numba \citep{lam2015numba}. We must support a range of versions, as the \textit{LiteBIRD} collaboration runs simulations on many HPC clusters, and each of them might support different versions of the Python interpreter. Table~\ref{tbl-version-numbers} shows the list of versions that have been officially released at the time of writing. When selecting version numbers, we follow the rules of semantic versioning. Until version 1.0 is released, we increment the second number whenever changes in the codebase introduce new features, and we increment the third number if the new release only contains bug fixes. (This has been the case with versions 0.2.1 and 0.15.1.)

\begin{longtable}[]{@{}ccc@{}}
\caption{List of LBS versions that have been officially
released.}\label{tbl-version-numbers}\tabularnewline
\toprule\noalign{}
Version & Release date & Supported Python versions \\
\midrule\noalign{}
\endfirsthead
\toprule\noalign{}
Version & Release date & Supported Python versions \\
\midrule\noalign{}
\endhead
\bottomrule\noalign{}
\endlastfoot
0.15.1 & June 2025 & 3.10-3.13 \\
0.15.0 & June 2025 & 3.10-3.13 \\
0.14.0 & February 2025 & 3.9-3.13 \\
0.13.0 & June 2024 & 3.9--3.12 \\
0.12.0 & March 2024 & 3.9--3.12 \\
0.11.0 & November 2023 & 3.9--3.12 \\
0.10.0 & June 2023 & 3.9--3.12 \\
0.9.0 & February 2023 & 3.7.1--3.9 \\
0.8.0 & October 2022 & 3.7.1--3.9 \\
0.7.0 & September 2022 & 3.7.1--3.9 \\
0.6.0 & July 2022 & 3.7.1--3.9 \\
0.5.0 & June 2022 & 3.7.1--3.9 \\
0.4.0 & December 2021 & 3.7.1--3.9 \\
0.3.0 & September 2021 & 3.6.1--3.8 \\
0.2.1 & March 2022 & 3.6--3.8 \\
0.2.0 & February 2022 & 3.6--3.8 \\
0.1.0 & September 2020 & 3.6-3.8 \\
\end{longtable}

In this paper, we describe version 0.11.0, which is the version used to implement the E2E script described in \cite{puglisi2024}.

We develop LBS on a public GitHub repository\footnote{\url{https://github.com/litebird/litebird_sim}} and release it under an open-source license (GPL3\footnote{\url{https://www.gnu.org/licenses/gpl-3.0.en.html}}); we chose open-source solutions due to their advantages in terms of accessibility and cost. The documentation is hosted publicly\footnote{\url{https://litebird-sim.readthedocs.io/en/master/index.html}}, and the manual is \emph{complete}: the policy of the development team is to merge contributions only if they contain appropriate additions/modifications to the User's manual and all the public functions and classes have their own docstrings. Moreover, the directory \texttt{notebooks}\footnote{\url{https://github.com/litebird/litebird_sim/tree/master/notebooks}} contains several Jupyter notebooks that show how to use the library to develop realistic pipelines.

We implemented LBS in Python, striking a balance between development efficiency and performance, as it offered a familiar and widely used language within the \textit{LiteBIRD} collaboration. While other choices, such as C++, Fortran, or Julia, might offer some performance advantages, Python has so far enabled faster prototyping and broader accessibility across the team, while providing adequate performance. Our code heavily uses NumPy \citep{harris2020array}, to handle arrays and matrices, and employs Numba \citep{lam2015numba} for those most CPU-intensive tasks where we measured a distinct advantage over NumPy. Numba has proven crucial in maximizing the performance while keeping the amount of memory allocated by modules like the pointing simulator reasonable, as explained in Section~\ref{sec-numba}. Moreover, Numba makes code deployment straightforward, as one does not need to ensure the availability of a C/C++/Fortran compiler on the host system. We use AstroPy \citep{astropy:2013, astropy:2018, astropy:2022} to track time and perform coordinate conversions and Healpy \citep{Zonca2019} to handle sky maps in HEALPix format \citep{Gorski2005}.

Since LBS is a library and not an executable, the user willing to run a simulation must first write a script that uses the library to perform the computations. Importing the library is easy, as it is listed on the Python Package Index\footnote{\url{https://pypi.org/project/litebird_sim}}; thus, LBS can be installed using the command

\begin{verbatim}
pip install litebird_sim==0.11.0
\end{verbatim}

\noindent where we specified version 0.11.0 because it is the subject of this paper. (Avoiding the version specification will install the most recent, which is 0.15.1 at the time of writing.)

\subsection{Memory layout}\label{sec-memory-layout}

LBS is designed to model the continuous data output of \textit{LiteBIRD}'s onboard instruments over its nominal three-year duration. LBS does not simulate low-level hardware functionalities such as the response of the optical system to its components (mirrors, struts, etc.) or the propagation of thermal instabilities within the mechanical structures. By focusing on the timelines of scientific samples rather than hardware-level operations, LBS tries to achieve a balance between simulation fidelity and computational feasibility.

The most significant feat in simulating the output of a three-year space mission with thousands of detectors is to allocate sufficient RAM for all the data structures. In this section, we describe the way the framework manages data.

\subsubsection{Scientific samples}\label{scientific-samples}

A key functionality provided by LBS is the allocation of the so-called Time-Ordered Data\footnote{LBS can also be used to run map-based simulations that do not need to simulate timelines. In fact, it has been used to develop map-based pipelines used internally by the \textit{LiteBIRD} collaboration. These simulations are orders of magnitude faster to run but produce less realistic results.} (TOD). A TOD is stored as a matrix where each row represents the simulated output timeline of a single detector. Storing this matrix requires substantial memory, potentially exceeding the capacity of typical computer systems when simulating multiple detectors at once: each detector samples the input signal from the telescope at a frequency of 19\,Hz using 32-bit floating-point numbers, thus requiring more than 6\,GB per detector.

To overcome potential memory limitations and accelerate calculations, LBS fully supports MPI for distributing the TODs across multiple processes. Considering a TOD matrix of shape \(N \times M\), where \(N\) is the number of detectors and \(M\) the number of samples, there are several ways to split it among \(k\) MPI processes:

\begin{itemize}
\item Splitting along the time axis results in each process holding a matrix of shape \(N \times (M / k)\). This layout is optimal when the simulation needs to work on the timelines of multiple detectors at once. For instance, noise correlations among detectors on the same wafer can be simulated efficiently using a correlation matrix and Cholesky decomposition. This data layout avoids inter-process communication overhead if no wafer hosts more than \(M / k\) processes.
\item Splitting along the detector axis leads to a TOD matrix with a \((N / k) \times M\) shape. This layout is typically employed when the simulation pipeline needs to compute Fourier transforms of the timelines.
\item More generally, one might want to split the number of detectors \(N\) and the number of time samples \(M\) simultaneously.
\end{itemize}

\begin{figure}
  \centering{
    \includegraphics[width=\textwidth]{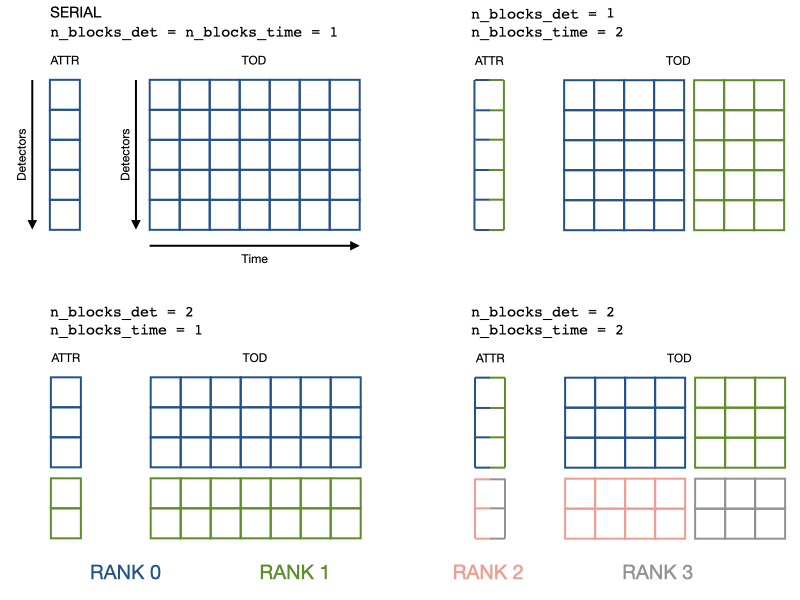}
  }
  \caption{\label{fig-data-distribution}The way LBS can split a 2D matrix containing a TOD, as well as any other attribute associated with each detector, like the noise level, the FWHM of the beam, etc. The first row illustrates how splits are made when the code is run in serial mode, i.e., without MPI. The parameters \texttt{n\_blocks\_det} and \texttt{n\_blocks\_time} specify the number of splits between the detectors and along the time axis, respectively. The bottom right panel shows how TODs are split when there are four MPI processes. Attributes can be copied over many MPI processes if these processes handle the same detectors. (The latter is the case of the two panels on the right.)}
\end{figure}

LBS lets the user to specify two parameters\footnote{Starting from version 0.14.0, LBS offers more sophisticated grouping options for \(N\), enabling detectors to be grouped based on their hosting wafer or other attributes. However, this feature was not used in \cite{puglisi2024} and will not be discussed further here.}: \texttt{n\_blocks\_det} and \texttt{n\_blocks\_time}, which define the number of splits for detectors and time samples, respectively. The default setting for both is 1, the only possible choice for serial applications where MPI is not employed. A visual depiction of the behavior of these two parameters is shown in Figure~\ref{fig-data-distribution}, which is taken from the LBS User's Manual.

Providing the ability to specify the split layout during the initial allocation of the TODs might not be enough for complex simulations that require running modules in sequence with different requirements. For instance, users might want to simulate the presence of correlations between detectors \emph{and} time correlations (\(1/f\) noise) in the same script: in this case, there is no optimal split layout that can work with both simulation modules. To accommodate these cases, LBS lets the user change the data splits \emph{after the TODs have been allocated}; this is done through the method \texttt{Observation.set\_n\_blocks()}, which accepts new values for \texttt{n\_blocks\_det} and \texttt{n\_blocks\_time} and reshuffles the samples in every TOD matrix across different MPI processes.

Another feature provided by LBS is the ability to work with multiple TOD matrices at once. When modeling data acquisition, it is often advisable to keep different signal components separate, such as the scientific signal and various noise contributions. Moreover, the sky signal itself typically consists of multiple astrophysical components, including the CMB, thermal emission from interstellar dust, and synchrotron emission from charged particles, among others. To track each component, LBS can create multiple 2D matrices per MPI process, ensuring that distinct contributions are represented individually, at the expense of increased memory usage.

Apart from the samples in a TOD, simulation modules often need additional information about the detectors being simulated. LBS can store an arbitrary number of attributes in memory; examples include the name of the wafer hosting the detector, the white noise level, the slope and knee frequency of the \(1/f\) noise. When the user configures LBS to distribute the \(N\) detectors across separate MPI processes, the framework ensures that the relevant attributes for each detector are also distributed to their corresponding process, enabling each process to have the necessary data for simulations and calculations available without duplication or the need for inter-process communication. For instance, if two processes \#1 and \#2 simulate four detectors A, B, C, and D, setting \texttt{n\_blocks\_det=2} and \texttt{n\_blocks\_time=1} will make the parameters for detectors A and B available only on process \#1. In contrast, the parameters for C and D will be present on process \#2. The method \texttt{Observation.set\_n\_blocks()}, which we described above and lets to change the split layout of the TOD matrices, automatically redistributes the attributes too.

\subsubsection{Pointing information and Half-Wave Plate
angles}\label{sec-pointing-information}

Beyond TODs, the collection of pointing information also requires consistent memory allocations. Pointing information specifies the direction and orientation of each detector's beam axis at the time when each sample was measured, and it is typically expressed through three angles: colatitude, longitude, and orientation (tilt). Consequently, the pointing matrix for \(N\) detectors and \(M\) time samples is a \(N \times M \times 3\) array. For MPI applications, LBS stores\footnote{As it is the case for CMB space surveys, the nominal scanning strategy is a composition of rotations with constant angular speed. In simulation codes, the typical approach is to encode these rotations as lowly-sampled quaternions, e.g., one per second or even less, depending on the angular speeds, and then use spherical linear operations (``slerp'') to compute pointing angles at the same sampling frequency as scientific data. Since version 0.13.0, LBS can compute pointings for each detector on the fly from quaternions, thus avoiding the need to store the pointing matrices and significantly reducing memory usage.} these matrices using the same distribution strategy as the TOD matrices.

In LBS 0.11.0, we used a simple scheme to encode the direction and orientation of the main beam of each detector as well as the angle of the HWP:

\begin{itemize}
\item The direction of the main beam was encoded by two angles (colatitude and longitude);
\item The orientation and the Half-Wave Plate angle were summed together and saved as an angle, called the ``polarization angle''.
\end{itemize}

The polarization angle is enough to simulate the behavior of a pencil beam and an ideal HWP, as under these assumptions, the map-maker has enough information to solve for I, Q, and U in each pixel. However, this approach provides insufficient information to simulate the systematic effects of non-ideal, asymmetric beams and realistic HWPs. For this reason, after version 0.11.0, we progressively implemented a more realistic model for pointings and HWP angles, which will be used in future simulation runs:

\begin{itemize}
\item Since version 0.12.0, LBS stores the HWP angle and the orientation as two distinct angles;
\item As the previous change increased the memory occupation, since version 0.13.0, pointings can be computed on-the-fly;
\item Starting from version 0.14.0, the polarization angle is no longer considered to be part of the orientation of the beam. This change is motivated by the integration of the \(4\pi\) beam convolution code provided by the Ducc\footnote{\url{https://gitlab.mpcdf.mpg.de/mtr/ducc}} library (see Section~\ref{sec-map-scanning}), as \(4\pi\) beams produced using electromagnetic simulation codes already encode the orientation of the polarization plane. (If no beam convolution is used in the code, it is still possible to tell LBS to add the polarization angle to the orientation angle when doing map-making, as it was the case before.)
\end{itemize}

\subsubsection{I/O}\label{io}

One of the purposes of LBS is to produce simulated data to be fed into data reduction pipelines, thereby validating the latter. To enable this interaction, LBS provides tools that save the timelines and the pointing information in HDF5 files. In version 0.11.0, LBS ensures that every MPI process saves its TODs and pointings in separate files; thus, a program running on \(N\) MPI processes will save \(2N\) files. To save disk space, the user can use a flag to instruct LBS to store the low-sampled quaternions used for computing the pointings instead of the full \(N\times M\times 3\) pointing matrix.

HDF5 metadata are used extensively to save ancillary information; examples include the list of names of the simulated detectors and their nominal sampling rates, their angular resolution, and a human-readable description of the simulation, among others.

Saving of HDF5 files has been crucial to interface LBS with the Commander pipeline, as explained in \cite{aurvik2024}.

\subsubsection{Other data}\label{sec-other-data}

Simulation modules often need to create new objects in memory, like sky maps or Fourier-transformed timelines. For MPI applications, modules usually allocate only those objects relevant to the data chunk hosted by their TOD matrix. A notable exception is the map-making module: as the maps are typically created from the samples acquired by several detectors and the resolution\footnote{The highest angular resolution of \textit{LiteBIRD} channels is 17.9\,arcmin at 402\,GHz, and the typical resolution of Healpix maps used in the \textit{LiteBIRD} collaboration is \texttt{NSIDE=512}, which leads to 36\,MB of storage space for a collection of three maps (I/Q/U).} of \textit{LiteBIRD}'s beams is not as high as for other CMB ground experiments, each MPI process receives a full copy of the final map.

\subsection{Provenance tracking}\label{sec-provenance-tracking}

Any simulation of a real scientific instrument needs several inputs that describe how the instrument is made and how it operates: what is the sampling frequency of the detectors, what are the characteristics of the noise profile, how many bits are used to measure one sample, what are the parts of the celestial sphere that are observed by the instruments during the nominal data acquisition, etc. This description is typically stored in a so-called Instrument Model (IMo), which is a database containing the details of each detector, optical system, hardware component, and so on. The \textit{LiteBIRD} collaboration has implemented the IMo in a versioned database, where each quantity is referred through a path and a version number. The implementation of the database is provided by InstrumentDB\footnote{\url{https://github.com/ziotom78/instrumentdb}.}, a Python program that manages information via a SQLite \citep{sqlite2020hipp} database. InstrumentDB is more than a plain database, as it provides a programmatic interface over the web that can be accessed by software written in any language. InstrumentDB is kept on a remote server hosted at the ASI--SSDC\footnote{\url{https://www.ssdc.asi.it/}} (Agenzia Spaziale Italiana Space Science Data Center), whose access is restricted both for its web interface, which \textit{LiteBIRD} members commonly access, and for its RESTful HTTP interface, which is usually used by scripts.

LBS can use the web interface to retrieve information from the database. However, as calls to a remote server can easily be a bottleneck for large-scale simulations that need a large number of inputs, InstrumentDB permits exporting the database to a folder, which can be copied to another computer and accessed locally by LBS itself. The latter is the preferred method for the LBS modules to fetch their inputs; in fact, this is the \emph{only} way\footnote{We purposefully prevented E2E scripts from querying the remote database to avoid saturating the network bandwidth at SSDC. Scripts based on LBS can access the remote database only if they send few requests per minute, otherwise the SSDC webserver will throttle the connection.} the E2E scripts described in \cite{puglisi2024} can operate.

The \textit{LiteBIRD} collaboration maintains a comprehensive description of the design of the spacecraft and of the instruments in the SSDC database server, which is restricted to members of the \textit{LiteBIRD} collaboration. However, the source code of LBS includes a reduced version of the database containing the basic design parameters that have already been published in \cite{Hazumi2021PTEP}. This allows users outside the collaboration to use LBS to run simulations, with the caveat that these may not represent the most up-to-date design of the experiment. For instance, the reference to the table containing general facts about the LFT instrument might be stored in the object whose path is

\begin{verbatim}
/releases/v10.3/satellite/LFT/instrument_info
\end{verbatim}

\noindent (assuming a hypothetical version 10.3 of the IMo), while the corresponding information taken from \cite{Hazumi2021PTEP} is stored in the publicly available object with path

\begin{verbatim}
/releases/vPTEP/satellite/LFT/instrument_info
\end{verbatim}

\begin{figure}
  \centering{
    \includegraphics[width=\textwidth]{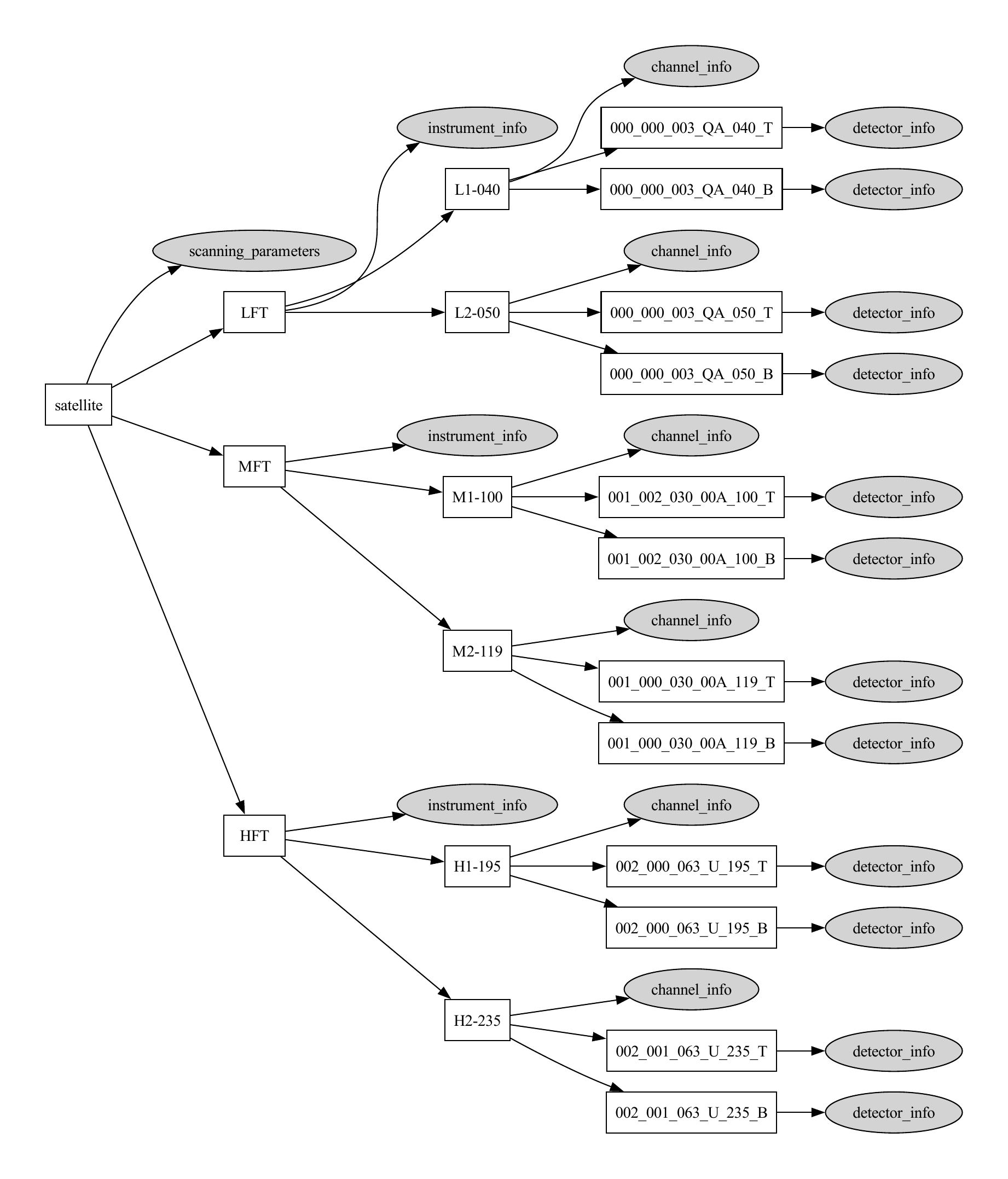}
  }
  \caption{\label{fig-ptep-db-tree}Simplified view of the PTEP IMo bundled with LBS 0.11.0. Boxes represent ``entities'', i.e., the nodes that provide the structure (topology) of the tree, while gray ovals represent the ``quantities'' that contain actual data. (A third level is represented by ``data files'', which are different versions of a quantity and are similar to commits in a Git repository.) The image only shows a few entities and quantities of the whole tree: for each instrument (LFT, MFT, HFT), only two channels are represented, and each channel shows only two detectors.}
\end{figure}

A simplified view of the public PTEP IMo bundled with LBS 0.11.0 is illustrated in Fig.~\ref{fig-ptep-db-tree}. The relevant information is stored in the so-called ``quantities'', which are represented as oval nodes with a gray background; all the quantities shown in the figure are JSON records containing several parameters that describe the ``entity'' to which they refer:

\begin{itemize}
\item A \verb|scanning_parameters| quantity contains the parameters of the nominal scanning strategy: spinning speed, angles of the rotation axes, etc.
\item A \verb|instrument_info| quantity contains general information about the instrument, such as the number of frequency channels or the angle between the boresight of the focal plane and the spin axis of the spacecraft.
\item A \verb|detector_info| quantity contains several synthetic parameters that describe the performance of a single detector, such as the NET and $1/f$ noise characteristics.
\item A \verb|channel_info| quantity contains the average values of several quantities associated with detectors operating at the same nominal frequency. They are the same as those found in the \verb|detector_info| objects, and are used for rough simulations where the detailed behavior of every single detector is not relevant.
\end{itemize}

As explained before, information is retrieved from the IMo through paths, which are plain Python strings. Thus, to retrieve the JSON dictionary associated with detector \verb|000_000_003_QA_040_T|, the path to be passed to LBS is

\vspace{0.5em}
\noindent \verb|satellite/LFT/L1-040/000_000_003_QA_040_T|
\vspace{0.5em}

Note that, as LBS refers to objects through their full path, it can handle any topology of the IMo. The official IMo database, which is not publicly available, uses a more complicated layout.

\section{Modules implemented in LBS}\label{sec-modules}

To simulate complex instruments like those onboard the \textit{LiteBIRD} spacecraft, multiple modules are required. Considering the task of simulating the nominal acquisition of the instruments, the following components are needed at a minimum:

\begin{itemize}
\item A flight simulator that models the way the spacecraft moves in space as time passes;
\item An optical simulator that considers the way photons propagate from celestial sources (CMB, ISM, point sources, \ldots) to the telescope;
\item An electronic simulator that simulates the way the signal captured by the optical system is measured and digitized by detectors;
\item A noise simulator that injects realistic instrumental noise in the samples measured by the detectors;
\item Etc.
\end{itemize}

We are not going to provide an exhaustive list of the modules implemented by LBS, as this list is still growing; the interested reader can refer to the User's manual available at \url{https://litebird-sim.readthedocs.io}, which includes the documentation of \emph{all} the modules\footnote{As stated in Section~\ref{sec-design}, the development team has the rule that no new module is integrated in the codebase if it is not fully documented in the User's Manual.} implemented in LBS.

In this section we will describe some of the modules implemented in LBS 0.11.0; we will then use these modules in Section~\ref{sec-full-example} to implement a simple pipeline.

For the sake of clarity, we group the modules in two sets: \emph{simulation modules} and \emph{data-reduction modules}. The formers have the task to simulate a process or the behaviour of some hardware, while the latters will be eventually replaced by the code that will be used to process the data acquired by the real instrument.

\subsection{Simulation modules}\label{simulation-modules}

\subsubsection{Simulation of input maps}\label{sec-sky-maps}

LBS provides the tools necessary to produce synthetic maps of the CMB and foreground sky at a specific observation frequency. These maps can then be observed through a simulation of the way the spacecraft spins, as discussed in more detail in Section~\ref{sec-scanning-strategy}. The production of synthetic sky maps is performed through the PySM3 \citep{Thorne2017, Zonca2021} library, which is conveniently wrapped by the \texttt{Mbs} module, whose acronym stands for Map-Based Simulation. Therefore, LBS supports all the models provided by PySM3.

\subsubsection{Scanning strategy}\label{scanning-strategy}

LBS can simulate the orbit of the spacecraft and compute the directions where each detector is looking at the sky as a function of time, i.e., the \emph{pointing information}. The \textit{LiteBIRD} spacecraft will scan the sky spinning around its spin axis while precessing around the Sun-Earth direction and orbiting around the Second Lagrangian point of the Sun-Earth system. The details of the scanning strategy are described in \cite{Hazumi2021PTEP}. Here, we provide some basic facts that help to understand how LBS produces the pointing information.

The most important thing to stress is that \textit{LiteBIRD} will perform a \emph{survey} of the sky, which means that the spacecraft will perform a set of periodic, continuous movements instead of pointing the telescope at several targets in sequence. The movements simulated by LBS are the composition of a set of rotations:

\begin{itemize}
\item The spacecraft rotates around its spin axis at a constant angular speed;
\item The spin axis of the spacecraft rotates around the Sun-Earth axis at a constant angular speed;
\item The Sun-Earth axis rotates around the Sun with a period of $~365$ days.
\end{itemize}

Of the three rotations, only the last one is not performed at a constant angular speed; instead, LBS uses AstroPy \citep{astropy:2022} to model the true revolution of the Earth around the Sun accurately\footnote{LBS can simulate a constant-speed revolution of the Earth around the Sun, for those simulations where accuracy is not as important as the speed of execution of the code. Moreover, the module can compute the expected velocity of the spacecraft with respect to the Sun. This can be used to simulate the kinetic dipole resulting from the spacecraft's motion with respect to the Sun. For this purpose, LBS implements an orbit simulator that simulates a Lissajous orbit, similar to the one followed by other spacecrafts orbiting \(L_2\) like WMAP and Planck.}. To accelerate pointing generation, the code encodes rotations using quaternions sampled at a tunable frequency that is typically lower than the detectors' sampling frequency, and then it employs a \emph{slerp} operation to compute the timelines of pointings at the nominal sampling rate.

\subsubsection{Instrumental noise}\label{instrumental-noise}

A realistic simulation of any instrument must include a noise component. Currently, LBS permits simulations of white noise and correlated noise with a \(1/f\) shape. There is no facility yet to introduce correlations in the noise between different detectors; however, this can be quickly implemented by multiplying the matrix containing the noise timelines by a proper correlation matrix. The power \(P(f)\) of the noise for each detector as a function of the frequency \(f\) is modeled by the following equation:
\begin{equation}
\label{eq:oneoverf}
P(f) = \sigma^2 \left(1 + \left(\frac{f_k}{f + f_\text{min}}\right)^\alpha\right),
\end{equation}
where \(\sigma^2\) is the power associated with white noise, \(f_k\) is the so-called \emph{knee frequency} of the correlated noise, \(\alpha\) is the \emph{slope} of the \(1/f\) component, and \(f_\text{min} \ll f_\text{samp}\) is a parameter that avoids the singularity at \(f = 0\). The noise simulation module provided by LBS requires the values of the three parameters \(\sigma^2\), \(f_k\), and \(\alpha\) to be provided for each detector.

Creating realistic \(1/f\) noise is technically challenging, as the amount of power associated with this component increases with the inverse of the frequency, and this means that there are correlations between samples separated by long time intervals. However, this prevents MPI-based simulations from splitting TODs along the time axis, unless some compromises are made on the correlation of the simulated noise timelines. There are noise generators that can produce noise with long correlations. Still, LBS 0.11.0 uses a simpler approach where white noise goes through a high-pass filter before being added to each \(N \times M\) matrix containing the scientific signal. This means that the noise coherence is not preserved across MPI processes. This approach has been employed by other CMB-oriented frameworks, such as \citep{toast:2021}, and has been shown to be sufficient for many applications, as the slowest noise terms are typically suppressed\footnote{Several techniques can be used to reduce low-frequency correlated noise, including relative gain calibration and map-making algorithms.} as part of the pre-processing. The problem of simulating \(1/f\) noise is a telling example of the difference between a mathematical model and a numerical model: even though the model expressed by \eqref{eq:oneoverf} is simple, a precise numerical implementation is not. We foresee that in the future we will probably need to implement a more sophisticated noise generator that avoids this problem.

\subsubsection{Solar dipole}\label{sec-solar-dipole}

LBS provides tools to simulate the signal associated with the relative velocity of the Solar System with respect to the CMB, i.e., the \emph{solar dipole}\footnote{LBS is able to simulate the kinematic dipole too. We leave the interested reader to the relevant chapter in the User's Manual: \url{https://litebird-sim.readthedocs.io/en/latest/dipole.html}.}. The CMB dipole is caused by a Doppler shift of the frequencies observed while looking at the CMB blackbody spectrum, according to the formula
\begin{equation}
  \label{eq:dipole}
   T(\vec\beta, \hat n) = \frac{T_0}{\gamma \bigl(1 - \vec\beta \cdot \hat n\bigr)},
 \end{equation}
where \(T_0\) is the temperature in the rest frame of the CMB, \(\vec \beta = \vec v / c\) is the dimensionless velocity vector, \(\hat n\) is the direction of the line of sight, and \(\gamma = \bigl(1 - \vec\beta \cdot \vec\beta\bigr)^2\).

CMB experiments usually employ the linear thermodynamic temperature definition, where temperature differences \(\Delta_1 T\) are related to the actual temperature difference \(\Delta T\) by the relation
\begin{equation}
   \Delta_1 T = \frac{T_0}{f(x)} \left(\frac{\mathrm{BB}(T_0 + \Delta T)}{\mathrm{BB}(T_0)} - 1\right) =
   \frac{T_0}{f(x)} \left(\frac{\exp x - 1}{\exp\left(x\frac{T_0}{T_0 + \Delta T}\right) - 1} - 1\right),
\end{equation}
where \(x = h \nu / k_B T_0\),
\begin{equation}
   f(x) = \frac{x e^x}{e^x - 1},
\end{equation}
and \(\mathrm{BB}(\nu, T)\) is the spectral radiance of a black-body according to Planck's law:
\begin{equation}
   \mathrm{BB}(\nu, T) = \frac{2h\nu^3}{c^2} \frac1{e^{h\nu/k_B T} - 1} = \frac{2h\nu^3}{c^2} \frac1{e^x - 1}.
 \end{equation}
 LBS implements several simplifications of Eq.~\eqref{eq:dipole} that we call ``dipole models''; they are reported\footnote{\url{https://litebird-sim.readthedocs.io/en/latest/dipole.html}} in the User's manual.

\subsubsection{Ideal HWPs}\label{sec-HWP}

\textit{LiteBIRD} implements Half-Wave Plates (HWPs) for each of its three instruments to rotate the polarization angle of the radiation entering the optical systems. In version 0.11.0, LBS simulates an ideal HWP, whose only effect is to alter the orientation angle of the detector, as described in Section~\ref{sec-pointing-information}. (More accurate simulators have been implemented in LBS 0.15.0.)

\subsection{Data-reduction modules}\label{sec-map-making}

Being a \emph{simulation} framework, LBS should not include data-reduction modules. Nevertheless, we implemented a few map-makers in LBS, as there are cases where maps are often more straightforward to analyze than timelines: their most significant advantage is that they take significantly less space and are easier to visualize.

The LBS map-makers save maps in FITS files using the Healpix\footnote{\url{http://healpix.sourceforge.net}} \citep{Zonca2019, Gorski2005} pixelization scheme. The following map-makers can be used with LBS:

\begin{itemize}
\item An internal binner, i.e., a simple map-maker that assumes that only uncorrelated noise is present in the timelines.
\item An internal destriper, i.e., a more advanced map-maker that can remove the effect of correlated instrumental noise from the timelines before producing a map.
\item A wrapper around the TOAST2 destriper \citep{toast:2021}, which is an optional dependency: this map-maker is available only if the user installed TOAST2 alongside LBS.
\item A function that exports the TOD to FITS files whose format is compatible with the Madam mapmaker \citep{keihanen:2005}.
\end{itemize}

\section{Validation}\label{sec-validation}

We describe here the problem of \emph{validation}, i.e., how to ensure that the pipeline produces correct outputs.

To validate a simulation software, it is necessary to ensure that the following requirements are met:

\begin{itemize}
\item The software must be \emph{accurate}, i.e., the results must match the expected within some reasonable threshold;
\item The software should be \emph{reliable}: if the input parameters are wrong, it should signal it, and in any case it should warn the user about unexpected features in the simulated data.
\item The results produced by the software should be \emph{reproducible}, i.e., anybody who has access to the source code and to the input parameters should be able to get the same outputs.
\item The code should be \emph{performant}, both in terms of speed and resource occupation (memory, disk space, etc.).
\end{itemize}

\subsection{Accuracy}\label{accuracy}

LBS is not a \emph{simulation software} but a framework. Thus, it is hard to determine whether its implementation is accurate enough or not, as an accuracy target is typically set for a specific pipeline. However, the modules in LBS can be validated so that their expected accuracy is enough for the kind of applications the framework is currently used in the collaboration. There are cases where we have implemented more than one algorithm for the same task: (1) we have two pointing generators that trade between speed and accuracy, and (2) we implemented several ways to produce a map from a set of timelines.

In LBS, we ensure that algorithms are accurate by means of several sets of automatic tests:

\begin{itemize}
\item \emph{Unit tests}, which test the correct behaviour and accuracy of single functions;
\item \emph{Integration tests}, which test the correctness and accuracy of single modules;
\item \emph{E2E tests}, which exercise multiple modules at once.
\end{itemize}

We discuss E2E tests in the companion paper \citep{puglisi2024}, so in this section we will describe only the unit tests and integration tests.

Unit tests ensure that single functions work correctly. They typically call the function to test with some easy-to-understand parameters and check that the result is what is expected. (These tests are good for documentation purposes as well.) The following example shows the tests for the function \texttt{compute\_pointing\_and\_polangle} (in \texttt{tests/test\_scanning.py}), used to compute the pointing direction and the polarization angle out of a quaternion representing the orientation of the boresight of a detector:

\begin{Shaded}
\begin{Highlighting}[]
\KeywordTok{def}\NormalTok{ test\_compute\_pointing\_and\_polangle():}
\NormalTok{    quat }\OperatorTok{=}\NormalTok{ np.array(lbs.quat\_rotation\_y(np.pi }\OperatorTok{/} \DecValTok{2}\NormalTok{))}
\NormalTok{    result }\OperatorTok{=}\NormalTok{ np.empty(}\DecValTok{3}\NormalTok{)}
\NormalTok{    lbs.compute\_pointing\_and\_polangle(result, quat)}
    \ControlFlowTok{assert}\NormalTok{ np.allclose(result, [np.pi }\OperatorTok{/} \DecValTok{2}\NormalTok{, }\FloatTok{0.0}\NormalTok{, }\OperatorTok{{-}}\NormalTok{np.pi }\OperatorTok{/} \DecValTok{2}\NormalTok{])}

    \CommentTok{\# We stay along the same pointing, but we\textquotesingle{}re rotating the detector}
    \CommentTok{\# by 90°, so the polarization angle is the only number that}
    \CommentTok{\# changes}
\NormalTok{    lbs.quat\_left\_multiply(quat, }\OperatorTok{*}\NormalTok{lbs.quat\_rotation\_x(np.pi }\OperatorTok{/} \DecValTok{4}\NormalTok{))}
\NormalTok{    lbs.compute\_pointing\_and\_polangle(result, quat)}
    \ControlFlowTok{assert}\NormalTok{ np.allclose(result, [np.pi }\OperatorTok{/} \DecValTok{2}\NormalTok{, }\FloatTok{0.0}\NormalTok{, }\OperatorTok{{-}}\NormalTok{np.pi }\OperatorTok{/} \DecValTok{4}\NormalTok{])}
\end{Highlighting}
\end{Shaded}

One of the limitations of unit tests is that the set of test inputs is necessarily far smaller than the overall set of possible inputs. Therefore, singularities and catastrophic cancellations that appear for non-trivial inputs can go undetected. A notable example in literature is \cite{duff:2017}, which discovered that in specific cases, the geometric algorithm implemented in \cite{frisvad:2012} produced incorrect results due to catastrophic cancellation; this error went undetected because, on average, this cancellation produced detectably wrong results only occasionally, and this is the reason why the problem went unnoticed in the original paper. An effective way to detect this kind of errors is to implement random testing, which calls the same function repeatedly over many random inputs; these tests are effective when verifying result correctness is straightforward (as in the case explained by \cite{frisvad:2012}). In LBS, an example is found in \texttt{tests/test\_mapmaking.py}, where the function \texttt{test\_cholesky\_and\_solve\_random()} tests that the function \texttt{solve\_cholesky} returns the solution of an equation of the form \(A x = v\), where \(A\) is a \(3\times 3\) Cholesky matrix\footnote{Cholesky matrices are used by the internal destriper to solve for the three Stokes components \(I\), \(Q\), and \(U\) of each pixel. Cholesky-based algorithms are advantageous in this context due to their low memory storage requirements, robustness, and high performance. We implemented a dedicated Cholesky solver tailored for \(3\times 3\) matrices, rather than using existing libraries, because we store the coefficients of the symmetric matrix in a custom data type that keeps the six nontrivial matrix coefficients in a single vector. Numba unrolls most of the loops in our code, and the performance of this routine is roughly 30\,\% more efficient than SciPy's \texttt{cholesky} function.} stored in a custom format used by the map-makers. The test compares the result with the solution of \texttt{numpy.linalg.solve} for a large number (1000 for LBS 0.11.0) of random Cholesky matrices.

Of course, unit tests are not enough to ensure that a simulation code is accurate and reliable, because errors can occur when combining several low-level functions into larger blocks. Integration tests verify that the simulation modules implemented in LBS work as expected by comparing the results expected by the analytical model with the actual results of the code. An example is found in \texttt{test/test\_destriper.py}, which tests that the results of the destriping map-maker are accurate. The destriper implements the model presented in \cite{kurki-suonio:2009}, where the destriping operation is represented through a linear operator that can be written in matrix form; however, the actual code avoids to build these matrices as they get huge in typical computations, so the numerical implementation of these equations must rely on more compact data structures. The tests implemented for the LBS destriper consider a minimal dataset consisting of only seven samples and observing a sky map of just 2 pixels; this case can be written in full matrix notation, and the test code checks that the results of the destriper with these inputs match the result obtained by simply inverting the destriping matrix. The test code is pretty long, as it amounts to roughly 1000 lines of code, because it includes unit tests, integration tests, and the code that builds the theoretical matrices and invert them.

\subsection{Reliability}\label{sec-reliability}

We have implemented several tests to ensure that the framework's foundations behave as expected in various situations. Dedicated tests for MPI processes are implemented in a separate script (\texttt{test/test\_mpi.py}), which checks the consistency of TOD/pointing matrices and attributes when different split layouts are used (see Section~\ref{sec-memory-layout}). These tests also verify the correct behavior of MPI-aware modules, such as the binner and the destriper.

In addition to tests, we have adopted a defensive approach in coding LBS and implemented several \texttt{asserts} in the code. These checks ensure the consistency of the parameters at each stage of the processing.

We list here only a few checks\footnote{At the time of writing, there are 128 assertions in the code, plus 63 \texttt{raise} statements that provide detailed messages when LBS detects problems in the input.} that the LBS modules perform:

\begin{itemize}
\item Incorrect splitting of the 2D matrix of samples across the MPI processes;
\item Incorrect correspondence between the components of a TOD and the components to be assembled by the map-maker;
\item Wrong order in the calls to modules (for instance, the removal of baselines from a TOD is called before destriping);
\item The caller requests a frequency outside the range of a detector bandpass;
\item Inconsistent inputs to the module that simulates the scanning strategy.
\end{itemize}

\subsection{Reproducibility}\label{sec-ensuring-reproducibility}

Apart from accuracy and reliability, simulation codes must ensure that their results are reproducible. There are several reasons why a software code called more than once produces different outputs:

\begin{itemize}
\item The input parameters might be different\footnote{This happens typically when the program asks the user to type input parameters using an interactive prompt. LBS does not provide any facility to input data in this way.} between different runs;
\item The person who ran the code might have been unaware that the source code was modified between two runs;
\item The code might depend on some hidden internal state, e.g., the presence of a file that has been updated between two runs, or the seed for the random number generator being initialized using the computer's clock.
\end{itemize}

It is generally impossible to ensure that none of these conditions occur, but LBS tries to minimize the likelihood of their happening.

First, when LBS is used in a script, it always generates a report in Markdown/HTML format in the output directory. This report includes several information that are useful for archival purposes:

\begin{itemize}
\item The date when the simulation was run;
\item The version number of LBS;
\item The hash of the most recent Git commit and the output of \texttt{git\ diff} (see below);
\item Other information.
\end{itemize}

The fact that the report is saved together with the simulation outputs (raw timelines, maps, plots, etc.) ensures that these outputs are easy to reproduce for several reasons that we detail here.

LBS encourages people implementing pipelines to provide the input parameters of their simulation scripts as parameter files in TOML format. (See Sect.\,Section~\ref{sec-provenance-tracking}.) These files are always copied to the output directory, and thus they can be passed as inputs to a new run of the simulation scripts.

Of course, ensuring that input files are the same is not enough, because the \emph{code} of the simulation program must be the same. It is often the case that people modify the code on the fly once they discover a bug but do not bump the version number! LBS provides a way to check source code consistency was not changed by assuming that the source code of the script that calls LBS is kept in a Git repository. LBS automatically saves the hash of the latest Git commit in the report saved in the output directory, allowing the user to verify whether two simulations were executed using the same commit or not. Unfortunately, it is often the case that programmers test the code before saving it in a commit; for this reason, if LBS detects that there are unsaved modifications in the code, it saves the output of \texttt{git\ diff} in the report as well.

Finally, there is the possibility that the code has some state that is not preserved across separate runs. There is no reliable solution that can work 100\% of the time to prevent hidden state\footnote{In principle, using functional languages like Haskell or Clojure and purely functional data structures might prevent this kind of error. However, no language is perfectly functional and thus the problem would still be unsolved.} from altering the results of a computation. However, a common source of confusion is the seed used to generate pseudo-random numbers, as it is often initialized to a value derived from the system clock. LBS requires that the parameter \texttt{random\_seed} accepted by the constructor of the class \texttt{Simulation} be \emph{always} provided. The seed can be \texttt{None} or an integer number; setting it to \texttt{None} uses the internal clock and thus prevents the reproducibility of the results; however, we feel the fact that \texttt{random\_seed=None} must be spelled explicitly in the source code to enable this behaviour is a good-enough ``red flag''. Instead, by setting \texttt{random\_seed} to an \texttt{int}, the results of separate runs of a script will produce the same sequence of pseudo-random numbers. The seed can be changed by using \texttt{sim.init\_random}, i.e., the same function used at the end of the \texttt{Simulation} constructor to set up the RNG:

\begin{Shaded}
\begin{Highlighting}[]
\NormalTok{sim.init\_random(}
\NormalTok{    random\_seed}\OperatorTok{=}\DecValTok{6789}\NormalTok{,}
\NormalTok{)}
\end{Highlighting}
\end{Shaded}

When running a parallel script, the \texttt{Simulation} constructor will take care of providing each process with a (different) dedicated RNG that produces uncorrelated sequences. The results will be reproducible when running the same script using the same \texttt{random\_seed} \emph{and} the same number of MPI processes. If the user uses the same seed but a different number of processes, the results will be different; however, LBS will include a warning in the output report.

\subsection{Performance}\label{performance}

In the development of LBS, we have verified that existing codebases like TOAST2, the simulation pipelines used for Planck, and other map-makers have performance similar to pipelines developed in our implementation. We also monitored the memory usage and speed of the E2E script described in \cite{puglisi2024} and collaborated closely with its development team to optimize memory usage and reduce CPU time.

As Python is notoriously slow when implementing loops, we have utilized NumPy and Numba to accelerate execution. Numba is particularly well-suited for loops that iterate over large arrays. It can easily beat NumPy, as the broadcast operation implemented by the latter requires multiple loops. We show an example in Section~\ref{sec-numba}.

\section{A full example}\label{sec-full-example}

So far, the description of the LBS framework has been only theoretical. To better highlight the features of the code, in this section we implement a simple E2E pipeline step-by-step. The reader can test all the code snippets in this section, provided that LBS 0.11.0 has been installed\footnote{The reader is advised to create a virtual environment before installing the \texttt{litebird\_sim} package and trying the commands listed in this work.} using \texttt{pip} and that the commands are executed in the same sequence they appear in this paper.

This example pipeline is not as polished as the official one described in the companion paper \citep{puglisi2024}. Here, our aim is to demonstrate the features of LBS and how its modules work together. The example will be split into several fragments that are meant to be read in the same order as they are presented in the paper. The interested reader can test the code on their computer, as the example was designed to be runnable on personal computers. We print the source code using colors to highlight the syntax, and if there is some output, we report it in black color under the code, like in the following example:

\begin{Shaded}
\begin{Highlighting}[]
\ImportTok{import}\NormalTok{ litebird\_sim }\ImportTok{as}\NormalTok{ lbs}
\BuiltInTok{print}\NormalTok{(}
  \SpecialStringTok{f"litebird\_sim }\SpecialCharTok{\{}\NormalTok{lbs}\SpecialCharTok{.}\NormalTok{\_\_version\_\_}\SpecialCharTok{\}}\SpecialStringTok{"}
\NormalTok{)}
\end{Highlighting}
\end{Shaded}

\begin{verbatim}
litebird_sim 0.11.0
\end{verbatim}

The code imports the LBS Python package, which is published on the Python Package Index under the name \texttt{litebird\_sim}, and it prints the version number. In the code fragments we are going to present in the next sections, we always assume that the Python package has been imported under the name \texttt{lbs}.

\subsection{Setting up the simulation}\label{setting-up-the-simulation}

The center point of LBS is the \texttt{Simulation} class, which should be instantiated in any pipeline built using this framework. The class serves as a container for several analysis modules available to the user, and it tracks both the inputs provided by the user and the output data generated by the simulation itself. Several pieces of information about the simulations need to be shared among different modules: for instance, the launch date and the duration of the simulation impact both the code that allocates the TOD, because it needs to know how many samples to allocate in memory, and the code that simulates the presence of moving sources in the sky, because it is required to compute the ephemerides of the planets.

Here, we show how to set up a simulation that lasts one day and starts on January 1st, 2024. We provide both a name and a description, which will be included in the report that is generated automatically at the end of any simulation:

\begin{Shaded}
\begin{Highlighting}[]
\ImportTok{import}\NormalTok{ astropy}

\NormalTok{sim }\OperatorTok{=}\NormalTok{ lbs.Simulation(}
\NormalTok{    base\_path}\OperatorTok{=}\StringTok{"./example"}\NormalTok{,}
\NormalTok{    start\_time}\OperatorTok{=}\NormalTok{astropy.time.Time(}
        \StringTok{"2024{-}01{-}01T00:00:00"}\NormalTok{,}
\NormalTok{    ),}
\NormalTok{    duration\_s}\OperatorTok{=}\FloatTok{86\_400.0}\NormalTok{,}
\NormalTok{    name}\OperatorTok{=}\StringTok{"My simulation"}\NormalTok{,}
\NormalTok{    description}\OperatorTok{=}\StringTok{"My description"}\NormalTok{,}
\NormalTok{    random\_seed}\OperatorTok{=}\DecValTok{12345}\NormalTok{,}
\NormalTok{    imo}\OperatorTok{=}\NormalTok{lbs.Imo(}
\NormalTok{        flatfile\_location}\OperatorTok{=}\NormalTok{lbs.PTEP\_IMO\_LOCATION,}
\NormalTok{    ),}
\NormalTok{)}
\end{Highlighting}
\end{Shaded}

The exact meaning of each keyword is explained in the User's Manual\footnote{\url{https://litebird-sim.readthedocs.io/en/latest/}.}; here we highlight a few points:

\begin{itemize}
\item Times are tracked using AstroPy\footnote{\url{https://www.astropy.org/}.}.
\item The parameter \texttt{duration\_s}, which takes the length of the simulation, shows a feature that is used extensively in the code: each quantity associated with a measurement unit reports the unit itself as part of the name. (In this case, the time must be expressed in seconds, hence the trailing ``\texttt{\_s}''.) This makes the code easy to read and reduces the chance of making conversion errors.
\item The \texttt{random\_seed} parameter initializes an internal pseudo-random number generator based on PCG-64\footnote{\url{https://numpy.org/doc/stable/reference/random/bit_generators/pcg64.html}.}. If the program is distributed using MPI, the \texttt{Simulation} class properly creates independent pseudo-random generators for each MPI process starting from this seed.
\item The \texttt{imo} parameter specifies where to look for the characteristics of the instrument. The constant \texttt{PTEP\_IMO\_LOCATION} refers to the data file containing a synthetic description of the instruments as provided in \cite{Hazumi2021PTEP}. This is not the official IMo for \textit{LiteBIRD}, but it has the advantage that it can be used freely, even by people who are not part of the collaboration.
\end{itemize}

The \texttt{imo} parameter is related to the way LBS tracks the provenance of the inputs for simulations, and its meaning will be explained in the next section.

\subsection{Accessing the IMo}\label{accessing-the-imo}

The constant \texttt{lbs.PTEP\_IMO\_LOCATION} in the code fragment shown in the previous section tells LBS that we will use the reduced IMo database in this example; in this way, any reader can run the code fragments in this paper, even if they do not have access to the (restricted) full \textit{LiteBIRD} IMo database.

Here is the code needed to access information about LFT, one of its frequency channels (40\,GHz), and two of its detectors:

\begin{Shaded}
\begin{Highlighting}[]
\CommentTok{\# Get a general description of the LFT instrument}
\CommentTok{\# (use the specification from the PTEP 2022 paper)}
\NormalTok{lft\_file }\OperatorTok{=}\NormalTok{ sim.imo.query(}\StringTok{"/releases/vPTEP/satellite/LFT/instrument\_info"}\NormalTok{)}
\BuiltInTok{print}\NormalTok{(}
    \StringTok{"The instrument }\SpecialCharTok{\{name\}}\StringTok{ has }\SpecialCharTok{\{num\}}\StringTok{ channels."}\NormalTok{.}\BuiltInTok{format}\NormalTok{(}
\NormalTok{        name}\OperatorTok{=}\NormalTok{lft\_file.metadata[}\StringTok{"name"}\NormalTok{],}
\NormalTok{        num}\OperatorTok{=}\NormalTok{lft\_file.metadata[}\StringTok{"number\_of\_channels"}\NormalTok{],}
\NormalTok{    )}
\NormalTok{)}
\end{Highlighting}
\end{Shaded}

\begin{verbatim}
The instrument LFT has 12 channels.
\end{verbatim}

This short code fragment shows the simplest way to access information in the IMo. The \texttt{sim.imo} object is an instance of the \texttt{Imo} class, which represents either a connection to a remote database or a link to a local copy of the IMo, as it is the case here. The \texttt{query} method accepts a path to a resource in the database; here, the odd-looking string \texttt{/releases/vPTEP/satellite/LFT/instrument\_info} uniquely indicates the kind of information we are accessing in the database. Specifically, the part \texttt{/releases/vPTEP} refers to the fact that we want a specific \emph{version} of the quantity: the one that was described in the so-called ``PTEP paper'', i.e., \cite{Hazumi2021PTEP}. (InstrumentDB can keep different versions of the same quantity, and in fact the restricted database used by the \textit{LiteBIRD} collaboration contains a more updated version of this quantity.) The remainder of the path-like string indicates that we are looking for a quantity named \texttt{instrument\_info}, which is stored within a pseudo-folder named \texttt{LFT}. The result is an object that provides several fields through its \texttt{metadatata} component; in the code fragment, we print the name of the instrument, ``LFT'' (kept under the key \texttt{name}) and the number of detector channels, 12 (kept under the key \texttt{number\_of\_channels}).

Importing quantities from the \texttt{metadata} field requires knowing their names, such as \texttt{name} or \texttt{number\_of\_channels} in the example above, which can make programming with LBS more challenging. For this reason, LBS provided a number of data classes like \texttt{InstrumentInfo}, \texttt{FreqChannelInfo}, \texttt{DetectorInfo}, etc., that allow writing more readable code. For instance, the code above can be rewritten using the class \texttt{InstrumentInfo} in the following way:

\begin{Shaded}
\begin{Highlighting}[]
\CommentTok{\# Ask the InstrumentInfo class to decode the "metadata"}
\CommentTok{\# of the quantity for us}
\NormalTok{lft\_instrument }\OperatorTok{=}\NormalTok{ lbs.InstrumentInfo.from\_imo(}
\NormalTok{    imo}\OperatorTok{=}\NormalTok{sim.imo, url}\OperatorTok{=}\StringTok{"/releases/vPTEP/satellite/LFT/instrument\_info"}
\NormalTok{)}

\CommentTok{\# Accessing the fields of "lft\_instrument" is done through}
\CommentTok{\# object fields, not Python dictionaries}
\BuiltInTok{print}\NormalTok{(}
    \StringTok{"The instrument }\SpecialCharTok{\{name\}}\StringTok{ has }\SpecialCharTok{\{num\}}\StringTok{ channels (again)."}\NormalTok{.}\BuiltInTok{format}\NormalTok{(}
\NormalTok{        name}\OperatorTok{=}\NormalTok{lft\_instrument.name,}
\NormalTok{        num}\OperatorTok{=}\NormalTok{lft\_instrument.number\_of\_channels,}
\NormalTok{    )}
\NormalTok{)}
\end{Highlighting}
\end{Shaded}

\begin{verbatim}
The instrument LFT has 12 channels (again).
\end{verbatim}

We have found that using these classes instead of directly accessing keys in the \texttt{metadata} dictionary has significantly improved our productivity, as modern IDEs and editors can suggest completions for field names. For example, when typing \texttt{name=lft\_instrument.{[}…{]}}, editors can trigger an auto-completion widget once the dot \texttt{.} has been typed and prompt for the available choices: \texttt{name}, \texttt{number\_of\_channels}, etc.

Every time we access the IMo, the \texttt{Simulation} object keeps track of our requests and stores them in a list that is saved alongside the output of the simulation. For instance, after the requests in the two code fragments above, the \texttt{sim} object contains the information that (1) we requested the \texttt{instrument\_info} quantity for LFT, and (2) we used the value of the quantity associated with the version named \texttt{vPTEP}, i.e., the description of the object that was presented in \cite{Hazumi2021PTEP}. This type of information is advantageous when archiving simulation outputs, as it clarifies what the inputs were and whether the results remain updated with the instrument's current design.

We are now going to fetch information about two LFT detectors that will be used in the simulation we are developing. As these detectors belong to LFT, we must first inform LBS that this is the instrument we will be simulating. In this way, LBS will know the proper orientation of the focal plane, as the orientations of the focal planes of LFT and MHFT are separated by 180° around the spin axis of the spacecraft. Setting the instrument is a matter of calling the method \texttt{Simulation.set\_instrument}:

\begin{Shaded}
\begin{Highlighting}[]
\NormalTok{sim.set\_instrument(lft\_instrument)}
\end{Highlighting}
\end{Shaded}

The next step is to load information about the detectors. This task is similar to what we have done above for the instrument, but this time we fill a \texttt{DetectorInfo} object instead of \texttt{InstrumentInfo}, and the pathlike string referring to the detector obviously changes. For the sake of simplicity and to keep the amount of calculations reasonable, we only load two detectors:

\begin{Shaded}
\begin{Highlighting}[]
\CommentTok{\# Get information about two 40 GHz detectors}
\NormalTok{det1 }\OperatorTok{=}\NormalTok{ lbs.DetectorInfo.from\_imo(}
\NormalTok{    imo}\OperatorTok{=}\NormalTok{sim.imo,}
\NormalTok{    url}\OperatorTok{=}\StringTok{"/releases/vPTEP/satellite/LFT/"}
\NormalTok{        }\StringTok{"L1{-}040/000\_000\_003\_QA\_040\_T/detector\_info"}\NormalTok{,}
\NormalTok{)}
\BuiltInTok{print}\NormalTok{(}
    \StringTok{"The NET of detector }\SpecialCharTok{\{det1\_name\}}\StringTok{ is }\SpecialCharTok{\{det1\_net\}}\StringTok{ µK·sqrt(s)"}\NormalTok{.}\BuiltInTok{format}\NormalTok{(}
\NormalTok{        det1\_name}\OperatorTok{=}\NormalTok{det1.name,}
\NormalTok{        det1\_net}\OperatorTok{=}\NormalTok{det1.net\_ukrts,}
\NormalTok{    )}
\NormalTok{)}

\NormalTok{det2 }\OperatorTok{=}\NormalTok{ lbs.DetectorInfo.from\_imo(}
\NormalTok{    imo}\OperatorTok{=}\NormalTok{sim.imo,}
\NormalTok{    url}\OperatorTok{=}\StringTok{"/releases/vPTEP/satellite/LFT/"}
\NormalTok{        }\StringTok{"LFT/L1{-}040/000\_000\_003\_QA\_040\_B/detector\_info"}\NormalTok{,}
\NormalTok{)}
\end{Highlighting}
\end{Shaded}

\begin{verbatim}
The NET of detector 000_000_003_QA_040_T is 114.63 µK·sqrt(s)
\end{verbatim}

In the examples shown so far, we have hard-coded the paths to detectors and objects in the Python code. However, in real-world applications, it is advisable to separate the code from the specification of the input data. LBS permits specifying the detectors to simulate using external parameter files saved in TOML\footnote{\url{https://toml.io/}.}. The following TOML file contains the same information that we provided in the code fragments presented so far:

\begin{Shaded}
\begin{Highlighting}[]
\NormalTok{[simulation]}
\NormalTok{base\_path = "./example"}
\NormalTok{start\_time = "2030{-}01{-}01T00:00:00"}
\NormalTok{duration\_s = 86400.0}
\NormalTok{name = "My simulation"}
\NormalTok{description = "My description"}
\NormalTok{random\_seed = 12345}

\NormalTok{[[detectors]]}
\NormalTok{detector\_info\_obj = "/releases/vPTEP/satellite/LFT/↩}
\NormalTok{    L1{-}040/000\_000\_003\_QA\_040\_T/detector\_info"}

\NormalTok{[[detectors]]}
\NormalTok{detector\_info\_obj = "/releases/vPTEP/satellite/LFT/↩}
\NormalTok{    L1{-}040/000\_000\_003\_QA\_040\_B/detector\_info"}
\end{Highlighting}
\end{Shaded}

This file can be passed to the constructor of the \texttt{Simulation} class instead of the parameters we used. This approach has a few advantages:

\begin{itemize}
\item All the parameters are kept in one file, and thus modifying the inputs is easier;
\item LBS copies the TOML file in the folder where the output of the simulation is saved, which can thus be run again, passing the same TOML file.
\end{itemize}

So far, we have loaded the descriptions of two detectors into memory; the next step is to instantiate the memory to store the samples measured by the simulated instruments. These are kept in ``observations'', which are the data structures that hold the timelines.

\subsection{Instantiating
observations}\label{sec-instantiating-observations}

LBS creates a set of \texttt{Observation} objects, which are basically wrappers around the 2D matrices containing the scientific samples see Sect.~\ref{sec-memory-layout}, and spreads them among the processes according to a scheme that can be tuned by the caller. Any \texttt{Observation} object can handle several 2D matrices at once; each of them is represented by a \texttt{TodDescription} object, where the acronym TOD stands for Time-Ordered Data, and it is common jargon in the CMB world for denoting these 2D matrices.

The way we allocate \texttt{Observation} objects is through the method \texttt{create\_observations} of the \texttt{Simulation} class. In the following example, we allocate two 2D matrices per each \texttt{Observation} object, where the first will contain the actual signal (the TOD), and the second will contain noise:

\begin{Shaded}
\begin{Highlighting}[]
\ImportTok{import}\NormalTok{ numpy }\ImportTok{as}\NormalTok{ np}
\NormalTok{sim.create\_observations(}
    \CommentTok{\# We are going to simulate the two detectors}
    \CommentTok{\# we fetched before}
\NormalTok{    detectors}\OperatorTok{=}\NormalTok{[det1, det2],}
    \CommentTok{\# For the sake of simplicity, here we just}
    \CommentTok{\# keep track of the sky signal and the noise;}
    \CommentTok{\# more realistic simulations would split "tod"}
    \CommentTok{\# into its components (CMB, dust, …)}
\NormalTok{    tods}\OperatorTok{=}\NormalTok{[}
\NormalTok{        lbs.TodDescription(}
\NormalTok{            name}\OperatorTok{=}\StringTok{"tod"}\NormalTok{,}
\NormalTok{            description}\OperatorTok{=}\StringTok{"TOD"}\NormalTok{,}
\NormalTok{            dtype}\OperatorTok{=}\NormalTok{np.float64,}
\NormalTok{        ),}
\NormalTok{        lbs.TodDescription(}
\NormalTok{            name}\OperatorTok{=}\StringTok{"noise"}\NormalTok{,}
\NormalTok{            description}\OperatorTok{=}\StringTok{"1/f+white noise"}\NormalTok{,}
\NormalTok{            dtype}\OperatorTok{=}\NormalTok{np.float32}
\NormalTok{        ),}
\NormalTok{    ],}
\NormalTok{)}
\end{Highlighting}
\end{Shaded}

\noindent (We use different datatypes for \texttt{tod} (64-bit floating point) and \texttt{noise} (32-bit floating point) for demonstration purpose.)

As we said above, if we ran our example using MPI, we could take advantage of the many processes by passing further arguments to \texttt{sim.create\_observation}; these arguments specify how the timelines should be split among the MPI processes. For the sake of simplicity, we assume that this example is ran serially using just one computer. However, more realistic codes can take advantage of several data split layouts, where the global 2D matrix is either split along rows (different MPI processes handle different detectors), along columns (different MPI processes simulate different time chunks), or both.

\subsection{Simulation of input maps}\label{simulation-of-input-maps}

Here we use \texttt{Mbs} to produce synthetic sky maps:

\begin{Shaded}
\begin{Highlighting}[]
\NormalTok{params }\OperatorTok{=}\NormalTok{ lbs.MbsParameters(}
    \CommentTok{\# Resolution of the maps to create}
\NormalTok{    nside}\OperatorTok{=}\DecValTok{32}\NormalTok{,}
    \CommentTok{\# Include the CMB}
\NormalTok{    make\_cmb}\OperatorTok{=}\VariableTok{True}\NormalTok{,}
    \CommentTok{\# Include the foregrounds}
\NormalTok{    make\_fg}\OperatorTok{=}\VariableTok{True}\NormalTok{,}
    \CommentTok{\# List of foregrounds: synchrotron (model \#0) and free–free (model \#1)}
\NormalTok{    fg\_models}\OperatorTok{=}\NormalTok{[}\StringTok{"pysm\_synch\_0"}\NormalTok{, }\StringTok{"pysm\_freefree\_1"}\NormalTok{],}
\NormalTok{)}

\CommentTok{\# We will simulate the sky as observed by the 40 GHz channels}
\NormalTok{channel }\OperatorTok{=}\NormalTok{ lbs.FreqChannelInfo.from\_imo(}
\NormalTok{    imo}\OperatorTok{=}\NormalTok{sim.imo,}
\NormalTok{    url}\OperatorTok{=}\StringTok{"/releases/vPTEP/satellite/LFT/L1{-}040/channel\_info"}\NormalTok{,}
\NormalTok{)}

\NormalTok{mbs }\OperatorTok{=}\NormalTok{ lbs.Mbs(}
\NormalTok{    simulation}\OperatorTok{=}\NormalTok{sim,}
\NormalTok{    parameters}\OperatorTok{=}\NormalTok{params,}
\NormalTok{    channel\_list}\OperatorTok{=}\NormalTok{[channel],}
\NormalTok{)}
\NormalTok{input\_maps }\OperatorTok{=}\NormalTok{ mbs.run\_all()[}\DecValTok{0}\NormalTok{]}
\end{Highlighting}
\end{Shaded}

The input maps are stored using the Healpix pixelization scheme \citep{Gorski2005}.

In the next section, we will describe how LBS simulates the scanning strategy and produces a simulation of the signal measurement from the synthetic sky stored in \texttt{input\_maps}.

\subsection{Scanning strategy}\label{sec-scanning-strategy}

The scanning strategy is encoded in the IMo via a set of angles and angular speeds, and thus it can be quickly retrieved using the IMo API. The following code loads the nominal scanning strategy described in \cite{Hazumi2021PTEP}, computes the quaternions, and produces the pointing information, which is stored in the same \texttt{Observation} objects that were allocated by the call to \texttt{sim.create\_observations} (see above).

\begin{Shaded}
\begin{Highlighting}[]
\NormalTok{sim.set\_scanning\_strategy(}
\NormalTok{    lbs.SpinningScanningStrategy.from\_imo(}
\NormalTok{        imo}\OperatorTok{=}\NormalTok{sim.imo, url}\OperatorTok{=}\StringTok{"/releases/vPTEP/satellite/scanning\_parameters"}
\NormalTok{    )}
\NormalTok{)}

\NormalTok{sim.compute\_pointings()}
\end{Highlighting}
\end{Shaded}

Each pointing is encoded as a set of three angles: the colatitude and longitude in Ecliptic coordinates (in radians), and the orientation angle of the detector projected in the same coordinate system; the latter is used to determine which linear polarization component is measured by the detector, thereby enabling the reconstruction of the Stokes $Q$ and $U$ parameters. It is possible to compute other parameters related to the spacecraft's motion, but we will discuss them in Section~\ref{sec-dipole}, where we will describe how LBS simulates the CMB dipole signal.

\subsection{An HWP}\label{an-hwp}

We include an ideal HWP using \texttt{IdealHWP}, which is a descendant of the class \texttt{HWP}:

\begin{Shaded}
\begin{Highlighting}[]
\NormalTok{sim.set\_hwp(}
    \CommentTok{\# The number is arbitrary}
\NormalTok{    lbs.IdealHWP(ang\_speed\_radpsec}\OperatorTok{=}\FloatTok{1.2345}\NormalTok{),}
\NormalTok{)}
\end{Highlighting}
\end{Shaded}

\subsection{Map scanning}\label{sec-map-scanning}

Once realistic sky maps have been produced (Section~\ref{sec-sky-maps}) and a scanning strategy is set (Section~\ref{sec-scanning-strategy}), it is possible to simulate the actual observation of the signal coming from the celestial sphere and detected by the bolometers once it has been focused by the optical system. At the moment, LBS is not able to simulate a convolution of the beam pattern of the optical system with the sky map, so the code assumes that the optical system is represented by perfect pencil beams\footnote{In LBS 0.15.0, we used a module in the Ducc library, called \texttt{totalconvolve}, that implements a high-speed algorithm for convolutions of functions over the sphere. With LBS 0.15.0, we can properly simulate the optical response of full \(4\pi\) beam patterns.} with a Full Width Half Maximum close to zero and no sidelobes; the axis of the pencil beam is aligned with the main axis of each beam.

Under the assumption of a pencil beam, scanning a map is just a matter of projecting the boresight direction of each detector on the celestial map simulated by PySM (Section~\ref{sec-sky-maps}) and storing the value of the corresponding pixel on the map in each sample in the timeline:

\begin{Shaded}
\begin{Highlighting}[]
\NormalTok{lbs.scan\_map\_in\_observations(}
\NormalTok{    sim.observations[}\DecValTok{0}\NormalTok{],}
\NormalTok{    input\_maps,}
\NormalTok{    input\_map\_in\_galactic}\OperatorTok{=}\VariableTok{False}\NormalTok{,}
\NormalTok{    component}\OperatorTok{=}\StringTok{"tod"}\NormalTok{,  }\CommentTok{\# Save the measurements in the "tod" 2D matrix}
\NormalTok{    interpolation}\OperatorTok{=}\StringTok{""}\NormalTok{, }\CommentTok{\# Do not interpolate pixels}
\NormalTok{)}
\end{Highlighting}
\end{Shaded}

The method \texttt{scan\_map\_in\_observation} can optionally perform a linear interpolation on the value of the pixels on the map through the function \texttt{pixelfunc.get\_interp\_val}, implemented by Healpy.

\subsection{CMB dipole}\label{sec-dipole}

To calculate the dipole signal, LBS must simulate an additional piece of information: the spacecraft's orbit, position, and velocity. The implementation of this calculation takes into account both the revolution of the Earth around the Sun and the orbit of the spacecraft around the Second Lagrangean point of the Sun-Earth system; the latter can optionally simulate a Lissajous orbit, as it was the case for the WMAP and Planck spacecraft, but in this example we will not turn on this option.

The code that implements the calculation of the velocity of the spacecraft is the following:

\begin{Shaded}
\begin{Highlighting}[]
\NormalTok{orbit }\OperatorTok{=}\NormalTok{ lbs.SpacecraftOrbit(}
\NormalTok{    sim.observations[}\DecValTok{0}\NormalTok{].start\_time,}
\NormalTok{)}
\NormalTok{pos\_vel }\OperatorTok{=}\NormalTok{ lbs.spacecraft\_pos\_and\_vel(}
\NormalTok{    orbit,}
\NormalTok{    sim.observations,}
\NormalTok{    delta\_time\_s}\OperatorTok{=}\FloatTok{60.0}\NormalTok{,}
\NormalTok{)}
\end{Highlighting}
\end{Shaded}

The variable \texttt{orbit} contains all the details about the motion of the Earth with respect to the Sun and about the orbit of the spacecraft, while \texttt{pos\_vel} is a matrix that contains the position and velocity of the spacecraft computed every minute (the parameter \texttt{delta\_time\_s}) for the whole period covered by the observations in \texttt{sim.observations}.

Once the variable \texttt{pos\_vel} is ready, we can inject the dipole signal into the timelines with a few lines of code:

\begin{Shaded}
\begin{Highlighting}[]
\NormalTok{dipole\_type }\OperatorTok{=}\NormalTok{ lbs.DipoleType.TOTAL\_FROM\_LIN\_T}
\NormalTok{lbs.add\_dipole\_to\_observations(}
\NormalTok{    sim.observations,}
\NormalTok{    pos\_vel,}
\NormalTok{    dipole\_type}\OperatorTok{=}\NormalTok{dipole\_type,}
\NormalTok{    component}\OperatorTok{=}\StringTok{"tod"}\NormalTok{,}
\NormalTok{)}
\end{Highlighting}
\end{Shaded}

The dipole signal is \emph{added} to the timeline named \texttt{tod}, so the result of the scanning of the synthetic maps described in Section~\ref{sec-map-scanning} that was added in \texttt{tod} too is not overwritten. We use the model \texttt{TOTAL\_FROM\_LIN\_T} that is described in Section~\ref{sec-solar-dipole}.

\subsection{Noise generation}\label{sec-noise}

For the sake of simplicity, we will include white noise in our example, and we will save it into the matrix called \texttt{noise}:

\begin{Shaded}
\begin{Highlighting}[]
\NormalTok{lbs.noise.add\_noise\_to\_observations(}
\NormalTok{    sim.observations,}
\NormalTok{    noise\_type}\OperatorTok{=}\StringTok{"white"}\NormalTok{,}
\NormalTok{    component}\OperatorTok{=}\StringTok{"noise"}\NormalTok{,}
\NormalTok{    random}\OperatorTok{=}\NormalTok{sim.random,}
\NormalTok{)}
\end{Highlighting}
\end{Shaded}

We did not specify the value of the noise parameters \(\sigma^2\), \(f_k\), and \(\alpha\) used in \eqref{eq:oneoverf}, as these are available to the module through the \texttt{DetectorInfo} class, which was already provided to the \texttt{Observation} objects instanced by \texttt{sim.create\_observations}; see Section~\ref{sec-instantiating-observations}.

\subsection{Map-making}\label{map-making}

In our example, we will use the internal destriper provided by LBS:

\begin{Shaded}
\begin{Highlighting}[]
\NormalTok{result }\OperatorTok{=}\NormalTok{ lbs.make\_destriped\_map(}
\NormalTok{    nside}\OperatorTok{=}\DecValTok{32}\NormalTok{,}
\NormalTok{    obs}\OperatorTok{=}\NormalTok{sim.observations,}
\NormalTok{    params}\OperatorTok{=}\NormalTok{lbs.DestriperParameters(),}
\NormalTok{    components}\OperatorTok{=}\NormalTok{[}\StringTok{"tod"}\NormalTok{, }\StringTok{"noise"}\NormalTok{],}
\NormalTok{)}
\end{Highlighting}
\end{Shaded}

\begin{verbatim}
[2025-04-11 20:30:09,554 INFO MPI#0000] Destriper CG iteration 1/100, ↩
    stopping factor: 7.028e-08
[2025-04-11 20:30:09,714 INFO MPI#0000] Destriper CG iteration 2/100, ↩
    stopping factor: 1.870e-08
\end{verbatim}

The variable \texttt{result} is an instance of the class \texttt{DestriperResults}, and it includes the following quantities:

\begin{itemize}
\item The estimates of all the baselines of the \(1/f\) noise component;
\item The destriped I/Q/U maps;
\item The hit map, where each pixel contains an integer number corresponding to the number of samples in the TOD that have been projected onto that pixel;
\item The binned I/Q/U maps, computed under the assumption that all the \(1/f\) baselines were zero (this is mostly useful to see the effect of the destriper);
\item Details about the convergence of the conjugated gradient algorithm and the overall time spent to produce the maps.
\end{itemize}

\subsection{TOD and map saving}\label{tod-and-map-saving}

The last part of our example saves the data produced by the simulation to disk. We have adopted the HDF5 format to save the timelines to disk, and the API to save the data is straightforward to call:

\begin{Shaded}
\begin{Highlighting}[]
\CommentTok{\# Save the timelines to HDF5 files}
\NormalTok{lbs.io.write\_observations(sim)}
\end{Highlighting}
\end{Shaded}

The HDF5 files contain both the timelines of samples and the complete pointing information, as well as other details about how the computation was split among MPI processes and other ancillary information that are useful for archival purposes.

As the maps produced by the destriper (Section~\ref{sec-map-making}) have been produced using the Healpix pixelization scheme, LBS employs the FITS format to save them; this ensures maximum compatibility with other codes and experiments:

\begin{Shaded}
\begin{Highlighting}[]
\NormalTok{sim.write\_healpix\_map(}\StringTok{"hit\_map.fits"}\NormalTok{, result.hit\_map)}
\NormalTok{sim.write\_healpix\_map(}\StringTok{"binned\_map.fits"}\NormalTok{, result.binned\_map)}
\NormalTok{sim.write\_healpix\_map(}\StringTok{"destriped\_map.fits"}\NormalTok{, result.destriped\_map)}
\end{Highlighting}
\end{Shaded}

\begin{verbatim}
PosixPath("example/destriped_map.fits")
\end{verbatim}

The LBS repository at \url{https://github.com/litebird/litebird_sim} contains a folder named \texttt{notebooks}, which includes several Jupyter notebooks that showcase various characteristics of the framework.

\section{Conclusions}\label{conclusions}

In this paper, we have presented the \textit{LiteBIRD} Simulation Framework (LBS), a Python library used for simulating the operations of the three instruments (LFT, MFT, and HFT) onboard the \textit{LiteBIRD} spacecraft.

LBS has been developed to provide a user-friendly yet robust framework. It implements several features to enhance the accuracy, reliability, and reproducibility of the results. These include the integration of fast Python libraries like NumPy and Numba, automated report generation, source code tracking, and versioned access to the instrument model.

LBS has successfully been used to conduct E2E simulations of the nominal data acquisition, with detailed results outlined in the companion paper \citep{puglisi2024}.

\appendix

\section{Using Numba to optimize intensive
computations}\label{sec-numba}

In Section~\ref{sec-design}, we explained that LBS utilizes NumPy for most of the code, but employs Numba for the most CPU-intensive tasks. We have chosen to use Numba on a case-by-case basis, depending on whether the E2E scripts required excessive memory or CPU time to run. Numba provided significant performance boosts in modules like the scanning strategy simulator and the generation of dipole timelines.

To illustrate how Numba can accelerate numerical codes that operate on long vectors, we consider a simple example. We have three arrays \(a\), \(b\), and \(c\), each containing \(10^6\) elements, and we want to compute the value \(2a_i + \sin b_i / 3\cos c_i\). The following code implements both calculations, which lead to equally correct results, as the printed numbers are the same:

\begin{Shaded}
\begin{Highlighting}[]
\ImportTok{import}\NormalTok{ numpy }\ImportTok{as}\NormalTok{ np}
\ImportTok{from}\NormalTok{ numba }\ImportTok{import}\NormalTok{ njit, prange}

\KeywordTok{def}\NormalTok{ numpy\_calculation(a, b, c, result):}
\NormalTok{    result[:] }\OperatorTok{=} \DecValTok{2} \OperatorTok{*}\NormalTok{ a }\OperatorTok{+}\NormalTok{ np.sin(b) }\OperatorTok{/}\NormalTok{ (}\DecValTok{3} \OperatorTok{*}\NormalTok{ np.cos(c))}


\AttributeTok{@njit}\NormalTok{(parallel}\OperatorTok{=}\VariableTok{True}\NormalTok{)}
\KeywordTok{def}\NormalTok{ numba\_calculation(a, b, c, result):}
    \ControlFlowTok{for}\NormalTok{ i }\KeywordTok{in}\NormalTok{ prange(}\BuiltInTok{len}\NormalTok{(result)):}
\NormalTok{        result[i] }\OperatorTok{=} \DecValTok{2} \OperatorTok{*}\NormalTok{ a[i] }\OperatorTok{+}\NormalTok{ np.sin(b[i]) }\OperatorTok{/}\NormalTok{ (}\DecValTok{3} \OperatorTok{*}\NormalTok{ np.cos(c[i]))}


\NormalTok{N }\OperatorTok{=} \DecValTok{1\_000\_000}
\NormalTok{a }\OperatorTok{=}\NormalTok{ np.random.rand(N)}
\NormalTok{b }\OperatorTok{=}\NormalTok{ np.random.rand(N)}
\NormalTok{c }\OperatorTok{=}\NormalTok{ np.random.rand(N)}
\NormalTok{result }\OperatorTok{=}\NormalTok{ np.empty(N)}

\NormalTok{numpy\_calculation(a, b, c, result)}
\BuiltInTok{print}\NormalTok{(}\StringTok{"First items calculated by NumPy: "}\NormalTok{, result[}\DecValTok{0}\NormalTok{:}\DecValTok{3}\NormalTok{])}
\NormalTok{numba\_calculation(a, b, c, result)}
\BuiltInTok{print}\NormalTok{(}\StringTok{"First items calculated by Numba: "}\NormalTok{, result[}\DecValTok{0}\NormalTok{:}\DecValTok{3}\NormalTok{])}
\end{Highlighting}
\end{Shaded}

\begin{verbatim}
First items calculated by NumPy:  [1.19619723 1.30527138 0.2299876 ]
First items calculated by Numba:  [1.19619723 1.30527138 0.2299876 ]
\end{verbatim}

However, if we measure the time spent in the two functions \texttt{numpy\_calculation()} and \texttt{numba\_calculation()}, the latter is significantly faster:

\begin{Shaded}
\begin{Highlighting}[]
\ImportTok{from}\NormalTok{ timeit }\ImportTok{import}\NormalTok{ timeit}

\BuiltInTok{print}\NormalTok{(}\StringTok{"""NumPy: }\SpecialCharTok{\{:.4f\}}\StringTok{ s}
\StringTok{Numba: }\SpecialCharTok{\{:.4f\}}\StringTok{ s}
\StringTok{"""}\NormalTok{.}\BuiltInTok{format}\NormalTok{(}
\NormalTok{    timeit(}\KeywordTok{lambda}\NormalTok{: numpy\_calculation(a, b, c, result), number}\OperatorTok{=}\DecValTok{5}\NormalTok{),}
\NormalTok{    timeit(}\KeywordTok{lambda}\NormalTok{: numba\_calculation(a, b, c, result), number}\OperatorTok{=}\DecValTok{5}\NormalTok{)))}
\end{Highlighting}
\end{Shaded}

\begin{verbatim}
NumPy: 0.0712 s
Numba: 0.0101 s
\end{verbatim}

This result shows that Numba is roughly 6 times faster than NumPy. A critical detail that led to this result is that before calling \texttt{timeit} we called once \texttt{numba\_calculation}: this triggered the Numba compiler, which compiled a machine-code version of the routine. This time is spent only once but it can affect benchmarks: if we avoided using \texttt{print()} in the example before, the result of \texttt{timeit} on \texttt{numba\_calculation()} would have included the compilation time, and thus the difference would have been smaller. To fully exploit Numba's advantages, we only used it for code that is called repeatedly or where the amount of time spent in processing is significantly larger than the time needed for Numba to compile the function. Examples include pointing generation and estimation of the dipole signal on all samples of a TOD, among others.

\acknowledgments\label{acknowledgments}

%
This work is supported in Japan by ISAS/JAXA for Pre-Phase A2 studies, by the acceleration program of JAXA research and development directorate, by the World Premier International Research Center Initiative (WPI) of MEXT, by the JSPS Core-to-Core Program of A. Advanced Research Networks, and by JSPS KAKENHI Grant Numbers JP15H05891, JP17H01115, and JP17H01125.
The Canadian contribution is supported by the Canadian Space Agency.
The French \textit{LiteBIRD} phase A contribution is supported by the Centre National d’Etudes Spatiale (CNES), by the Centre National de la Recherche Scientifique (CNRS), and by the Commissariat à l’Energie Atomique (CEA).
The German participation in \textit{LiteBIRD} is supported in part by the Excellence Cluster ORIGINS, which is funded by the Deutsche Forschungsgemeinschaft (DFG, German Research Foundation) under Germany’s Excellence Strategy (Grant No. EXC-2094 - 390783311).
The Italian \textit{LiteBIRD} phase A contribution is supported by the Italian Space Agency (ASI Grants No. 2020-9-HH.0 and 2016-24-H.1-2018), the National Institute for Nuclear Physics (INFN) and the National Institute for Astrophysics (INAF).
Norwegian participation in \textit{LiteBIRD} is supported by the Research Council of Norway (Grant No. 263011 and 351037) and has received funding from the European Research Council (ERC) under the Horizon 2020 Research and Innovation Programme (Grant agreement No. 772253, 819478, and 101141621).
The Spanish \textit{LiteBIRD} phase A contribution is supported by MCIN/AEI/10.13039/501100011033, project refs. PID2019-110610RB-C21, PID2020-120514GB-I00, PID2022-139223OB-C21, PID2023-150398NB-I00 (funded also by European Union NextGenerationEU/PRTR), and by MCIN/CDTI ICTP20210008 (funded also by EU FEDER funds).
Funds that support contributions from Sweden come from the Swedish National Space Agency (SNSA/Rymdstyrelsen) and the Swedish Research Council (Reg. no. 2019-03959).
The UK  \textit{LiteBIRD} contribution is supported by the UK Space Agency under grant reference ST/Y006003/1 - "LiteBIRD UK: A major UK contribution to the LiteBIRD mission - Phase1 (March 25)."
The US contribution is supported by NASA grant no. 80NSSC18K0132.
%

We acknowledge the use of CINECA HPC resources from the \textit{LiteBIRD} INFN project and the computing facilities provided by NERSC.

We acknowledge the use of Google Gemini 2.0 and Grammarly to generate suggestions for improving writing style and proofreading. No content generated by AI technologies has been presented as our own work.

\section*{Bibliography}\label{bibliography}

\renewcommand{\bibsection}{}
\bibliography{bibliography.bib}

\end{document}